\documentclass[twocolumn]{aastex63}
\usepackage{graphicx}
\usepackage{txfonts}
\usepackage{setspace} 
\usepackage[inline, shortlabels]{enumitem}

\def\ie{{\em i.e., }}
\def\be{\begin{equation}}
\def\bea{\begin{eqnarray}}
\def\eea{\end{eqnarray}}
\def\ee{\end{equation}}
\def\gmin{\Gamma_{\rm min}}
\def\rmin{r_{\rm min}}
\def\zmin{z_{\rm min}}
\def\zmax{z_{\rm max}}
\def\Rph{R_{\rm ph}}
\def\Lj{L_{\rm j}}
\def\tj{\theta_{\rm j}}

\def\te{\theta_{\rm e}}
\def\mdj{\dot{M}_{\rm j}}
\def\mda{\dot{M}_{\rm a}}
\def\La{L_{\rm a}}
\def\g0{\Gamma_{0}}
\def\to{\theta_{\rm o}}
\def\Q{\cal{Q}}

\graphicspath{{./}{figures/}}

\begin{document}
\title{\large {Optically thick jet base and explanation of edge brightening in AGN jets}}
\setstretch{1.0}

\correspondingauthor{Mukesh Kumar Vyas} 
\email{mukeshkvys@gmail.com}


\author{Mukesh Kumar Vyas}
\author{Asaf Pe'er}
\affiliation{Bar Ilan University, \\ Ramat Gan, Israel,
 5290002}





\begin{abstract}
The jet cores in blazars are resolved and found to harbour an edge brightened structure where the jet base appears extended at sides compared to its propagation axis. This peculiar phenomenon invites various explanations. We show that the photosphere of an optically thick jet base in Active Galactic Nuclei (AGNs) is observed edge brightened if the jet Lorentz factor harbours an angular dependence. The jet assumes a higher Lorentz factor along the jet axis and decreases following a power law along its polar angle. For an observer near the jet axis, the jet has a lower optical depth along its propagation axis compared to off axis regions. Higher optical depths at the outer region makes the jet photosphere appear to extend to larger radii compared to a deeper photosphere along its propagation axis. We tackle the problem both analytically and numerically, confirming the edge brightening through Monte Carlo simulations. Other than the edge brightening, the outcomes are significant as they provide a unique tool to determine the jet structure and associated parameters by their resolved observed cores. The study paves way to explore the spectral properties of optically thick cores with structured Lorentz factors in the future.
\end{abstract}
\keywords{High energy astrophysics; Active Galactic Nucleus; Relativistic jets; Theoretical models}

\section{Introduction }
\label{sec_intro}
Blazars are a class of Active Galactic Nuclei (AGNs) where the observer is situated near the jet axis. Radio intereferomety enables a deeper view of the jet base  to reveal their structure at launching \citep{1978Natur.272..131R,1979ApJ...231..293C,1981Natur.290..365P,1981ApJ...248...61P}. The cores of extragalactic AGN jets are launched from inner regions of accretion discs within around 100 Schwarzschild radii \citep{1999Natur.401..891J,2012Sci...338..355D}. The jet core appears to have an extended structure towards its edges compared to its centre, a phenomenon called limb brightening \citep{2006evn..confE...2K, 2014evn..confE..13K, 2016Galax...4...39K, 2021Galax...9...58G}. Some examples include the event horizon telescope (EHT) image of Centaurus A, \citep[][]{janssen2021event}, M87 core with wide opening angle \citep[][]{2016Galax...4...46W} etc. Understanding this peculiar shape of the jet base is an intriguing problem.

The edge brightening in AGN jets has several explanations. \cite{2011MNRAS.415.2081C} attributed this phenomenon to skewness in synchrotron emission. Due to helical magnetic fields in the jet, the synchrotron emission is anisotropic and more prominent from off axis regions compared to the jet axis. Hence the outer boundary of the jet appears brighter when the observer is nearly along the axis. Similar argument applies for toroidal magnetic fields. In relativistic magnetohydrodynamic simulations, limb brightening is shown to be caused by the toroidal magnetic fields in jets launched by Blandford–Znajek mechanism 
\citep{2021A&A...656A.143K, 2018ApJ...868...82T}.
Additionally, the observed limb brightening at large distances from the jet base is reported to arise due to recollimation shocks such as seen in the narrow-line Seyfert galaxy $1H~0323+342$ \citep{2018ApJ...857L...6D}. 

Most common explanation to edge brightening is given by a spine sheath structure of the jet. In this model, a fast moving jet at the centre is surrounded by a slow moving wind like flow. Such two component jets were predicted theoretically  \citep{1989MNRAS.237..411S,1991ApJ...383L...7H,1996ASPC..100..241L,2003NewAR..47..667M}. Later they were seen in several numerical investigations using general relativistic magneto-hydrodynamic (GRMHD) simulations of extragalactic jets \citep{2006ApJ...641..103H, 2006MNRAS.368.1561M,2007ApJ...664...26H}. It was shown that a jet spine is formed along its propagation axis due to magnetic fields threading the ergosphere. This beamed spine is surrounded by a wider sheath generated and driven by the anchored magnetic fields in the accretion discs. 
The spine-sheath structure is used to explain limb brightening in AGN jets \citep{1990PAZh...16..661K, 1990SvAL...16..284K} such as Markarian 501 \citep[][]{2004ApJ...600..127G}, M87 \citep{2018A&A...616A.188K} and some radio galaxies \citep{1998ApJ...507L..29S,2001ApJ...552..508G}. The phenomenon is attributed to Doppler deboosting \citep{1990SvAL...16..284K}, where the inner region of the jet propagates faster leading to relativistic beaming of the photons along the direction of propagation. This makes a smaller photon flux observed by an off axis observer. The relativistic beaming is less in outer regions of the jet as it propagates slowly leading to a brighter limb compared to the jet axis. Hence, the spine-sheath model used to explain the limb brightening is due to the relativistic beaming from optically thin plasma. It predicts a higher observed intensity from the outer regions compared to inner region.

The spine-sheath model is invoked to explain several other observed phenomena in AGNs including TeV emission from blazars \citep{2005A&A...432..401G,2008MNRAS.385L..98T, 2008MNRAS.386L..28G}, efficient neutrino productions in AGNs \citep{2014ApJ...793L..18T, 2023arXiv230413044N} and a broadband emission from PKS 1127-145 \citep{2007ApJ...657..145S}. 

However, the conventional explanation to limb brightening by spine-sheath jet structure is sometimes debated and considered to be a strived explanation \citep{2021Galax...9...58G}. The explanation may not be sufficient to explain limb brightening in general for several reasons. Jets with persistent direction change do show limb brightening (example includes Mrk 501). When the jet swings, limb brightening is likely to disappear following the Doppler boosting explanation in spine sheath model \citep[see][for a review]{2021Galax...9...58G}. Additionally, sometimes the limb brightening is seen at large distances from the jet base \citep{2018ApJ...857L...6D} which requires an explanation beyond the Doppler beaming due to spine sheath model.


In recent years, in the images of resolved jet cores, bright surfaces indicating optically thick emission region is visible near the jet base \citep{2006evn..confE...2K, 2014evn..confE..13K, 2016Galax...4...39K}. Such a surface indicates the transition from optically thick to optically thin region, called a photosphere. The location of this photosphere is close to the central black hole (BH) when viewed close to the jet axis and extends to larger radii at larger polar angles, thereby having a concave shape \citep{2017A&ARv..25....4B, janssen2021event}.
Motivated by these observations of the photospheric surface, here we provide an alternate explanation to the limb brightening with an angle dependent Lorentz factor profile for the jet. In our explanation, it is caused by the emission from optically thick region in an angle dependent jet.

The optically thick region in AGN jets is due to synchrotron self absorption \citep{1979ApJ...232...34B,2008ChJAA...8...39M,2018A&A...612A..67G, 2022A&A...668A...3B}, and as we show here due to high particle density for Compton scattering as well. At parsec scale, the electron density in AGN jets is around few hundred to few thousand particles per cm$^{-3}.$ This value is inferred by various observational techniques such as core shift analysis \citep{1998A&A...330...79L}, Faraday rotation \citep{2021ApJ...910...35L} and spectral analysis of AGN jets \citep{2016ApJ...826..135L}. To estimate the density near the jet base, we can assume an inverse square variation of electron density with distances ($n\propto r^{-2}$). It is not only a natural representation of the density with distance, it is an observational requirement as well \citep{1981ApJ...243..700K,1998A&A...330...79L,2021ApJ...910...35L}. Thus, calculating particle density close to the jet base (few Schwarzschild radii) we show that the jet base should be optically thick for Compton scattering. Furthermore, the evidence of high mass loading in jets  \citep{2021ApJ...923..256Q} suggests the possibility of treating the jet core to be optically thick. As we show below, for plausible measured densities, the photospheric radius in many AGN jets extends much beyond the Schwarzschild radius.

For an on axis observer, a jet having a decreasing Lorentz factor with polar angle leads to an angle dependent optical depth along its angular extent. Due to higher Lorentz factor along the jet axis, the jet stem is optically thin and the photosphere appears deeper compared to optically thicker region at off axis \citep{1991ApJ...369..175A, 2008ApJ...682..463P, 2023ApJ...943L...3V}. This makes the photosphere extended at larger polar angles as compared to the jet axis, thus creating limb brightening in jets.

We develop a semi analytic model to estimate the photosphere in this framework. Additionally, we perform Monte Carlo simulations for photon scattering at the jet base and conclude that a jet with a structured Lorentz profile produces an edge brightened photosphere for all observers at small observing angles. Alternatively, the observed images of these jets allow us to infer the jet structure and paves way to investigate other spectral features of these jets in light of their photospheric emission. In section \ref{sec_calc} below, we formulate the theoretical analysis for estimating the jet photosphere as a function of coordinates. In section \ref{sec_results} we discuss the results obtained and conclude the study in section \ref{sec_conclusions}.



\section{Photosphere of a relativistic fluid with angle dependent Lorentz factor ($\Gamma$)}
\label{sec_calc}
\subsection{Existence of a Compton photosphere at AGN jet base}
\label{sec_sub_compton_phot}
We consider the AGN jets to be optically thick for Compton scattering at the base. 
The optical depth for scattering of a photon propagating a distance $dr$  by a fluid element with density $n$ 
is $d\tau = \sigma n dr$. Here, it is safe to take the Thomson cross section, namely $\sigma= \sigma_T$. For approximately constant outflow velocity and conical jets, one can assume an inverse square law for the particle density, $n(r) = n_0 (r_0/r)^{2}$. 
Integrating the optical depth from the jet base at $\approx 2 r_s$ (where $r_s$ is the Schwarzschild radius of the central BH) to the observer at infinity, the required condition for optically thick plasma is
\be 
\tau_b = \int d\tau \sim 10^3 \left(\frac{10^6}{m_{\rm BH}}\right)\times\left(\frac{n_0}{100}\right)\times \left(\frac{r_0}{1 {\rm parsec}}\right)^2>1.
\ee
Here, $m_{BH}$ is black hole mass in units of solar mass, and $n_0$ is given in units of particles/cm$^3$. Using observationally estimated values of $n_0$, $r_0$ and $m_{\rm BH}$, one can assert the optical depth in AGN jet bases. For example in the quasar 3C 273, having BH mass $m_{\rm BH} = 6.59 \times 10^{9}$ \citep{2005A&A...435..811P}, the estimated particle density at distance $7 \times 10^{20}$ cm is $125$ cm$^{-3}$ \citep{2021ApJ...910...35L}. The optical depth at the jet base is therefore $\tau_b = 10^4$. 
Similar analysis for quasar 3C 207 gives $\tau_b = 7.87$. [Density and distance are estimated by \citet{2004ApJ...608..698S}, while the black hole mass is estimated in \cite{2012ApJS..201...38T}]. For 3C 345 it turns out to be 5.63 [Density and distance are taken from \cite{2004ApJ...608..698S} while the black hole mass is estimated in \cite{2007ApJ...661..719U}]. Core shift analysis gives particle densities at around parsec scale to be 1500 cm$^{-3}$ \citep{1998A&A...330...79L} for various sources. This leads to a typical optical depth at the jet base in range $\tau_b \sim 3- 3\times 10^3$ for black hole masses $m_{\rm BH} = 10^6-10^9$. However, it is not the case with all the AGN jets and sometimes the core turns out to be optically thin. One such example is PKS $(1136-135)$, the estimated optical depth at the jet base is 0.21 [$n_0$ and $r_0$ estimated by \cite[][]{2004ApJ...608..698S} and the black hole mass is taken from 
\cite[][]{2007ApJ...661..719U}].  

It is important to mention that the density estimates in an AGN jet has a large uncertainty associated with the applied method. For example, the density estimates in M87 jet using rotation measure at distance $r_0 = 156$ parsec is $n_0 = 1.6 \times 10^{-3}$ cm$^{-3}$ giving $\tau_b = 0.06$. While from  X-ray energy spectral analysis, the density estimates come out to be four orders of magnitude greater leading to $\tau_b \sim 600$ \citep{2023arXiv230712039O}. 
In these estimates, the black hole mass of M87 is $6.5 \times 10^9$ solar mass, taken from \cite{2019ApJ...875L...6E}.
In the estimates above, the typical calculated particle densities close to the jet base, at the distance of a few Schwarzschild radii reach up to $10^{10-12}$ cm$^{-3}$ . Such high densities in AGN atmospheres are reported by various authors.
For example, the density inside the jet in 3C 84 is estimated at $10^3-10^5$ cm$^{-3}$ at $0.07-0.14$ parsec \citep{2021ApJ...920L..24K,2017ApJ...849...52N,2018ApJ...864..118K}. The intermediate line regions near AGN jets have densities $10^5$ cm$^{-3}$ to $10^{11.5}$ cm$^{-3}$ at characteristic distances of 0.1 parsec \citep{2017FrASS...4...19A,2016ApJ...831...68A}.
Particle densities in the region above AGN accretion discs are shown to be  $\approx 10^{15}$ cm$^{-3}$ \citep{2016ApJ...831...68A}.
Hence we conclude that the optically thick cores in AGN jets exist in many sources and therefore the photospheric properties of such cores need to be explored.

\subsection{Jet structure}
The existence of a spine-sheath structure in AGN jets has both theoretical and observational basis. 
Here we consider an angle-dependent jet velocity profile, $\Gamma = \Gamma(\theta)$. 
In order to avoid abrupt change in Lorentz factor between spine and sheath, we consider a smooth transition following a power law decay with polar angle $\theta$, given as
\be 
\Gamma(\theta)= \gmin+\frac{\Gamma_{0}-\Gamma_{\rm min}}{\sqrt{\left(\frac{\theta}{\theta_{\rm j}}\right)^{2p}+1}}.
\label{eq_gam_profile}
\ee 
Here $\g0$ and $\gmin$ are the maximum and minimum values of the jet Lorentz factor assigned to the spine and sheath regions respectively. The angle $\tj$ is a constant, separating the inner, faster jet core from the outer, slower sheath. The parameter $p$ is a jet profile index that determines the steepness of decrease in $\Gamma$ with $\theta$. This profile implies an inner jet region (at angles $\theta \ll \theta_j$) having constant Lorentz factor $\Gamma=\Gamma_0$ and outer jet region, at angles larger than  $\theta= \te = \theta_{\rm j}\Gamma_0^{1/p}$ has $\Gamma=\Gamma_{\rm min}$. Within the region $\tj-\te$, the Lorentz factor decays following a power law $\Gamma\propto \theta^{-p}$. This profile is analogous to the evolution of $\Gamma$ obtained by \cite{2006MNRAS.368.1561M} in their simulations [see their Figure 9, first panel]. This continuous transition of Lorentz factor between the spine and sheath arises from mutual interaction and mixing of particles in these regions. However, the conclusions of our study are independent of the form of the transition and are valid for a step function transition as well (obtained here for $p \rightarrow \infty$).

\subsection{Theoretical model : Estimation of optical depths in a structured jet}
Consider that the photons are emitted deep inside the jet. These photons escape once they reach the photospheric radius $\Rph$ where the optical depth for scattering between $\Rph$ and infinity equals unity. In such a case, the shape of $\Rph$ as a function of the polar coordinates ($\theta, \phi$) determines the appearance of the jet base in the observations. The observed shape is sensitive to the given observer's location, specified here by polar coordinates $\to$ and $\phi_{\rm o}$. 
We calculate here the angular dependence of $\Rph$ for a given 
observer's location. We calculate $\Rph$ both analytically and numerical simulations as follows.

Define a cylindrical coordinate system centered at the plasma expansion center (base of the jet) and assume that the observer is located at plus infinity on the $z$-axis. Since the observer is, in general, off the jet axis, the jet outflow is not symmetric in this coordinate system. Consider a photon that was emitted at point $z_{\rm min}$ along the $z$ axis and at radial distance $r_{\min}$  from it, namely at distance $r = \left(r_{\rm min}^2 + z_{\rm min}^2 \right)^{1/2}$ from the center. Assume that this photon propagates towards the observer (along the $+z$ direction). The optical depth as measured along the ray traveling in the $+z$ direction and reaching the observer is 
  \citep{1991ApJ...369..175A,2008ApJ...682..463P}, 
\be 
\tau(r_{\rm min}, z_{\rm min}) = \int_{\zmin}^{\infty} n' \sigma_{\rm T} \Gamma [1-\beta \cos \theta_l]dz.
\label{eq_tau1}
\ee 
Here, $n'$ is the electron number density in the local comoving frame, $\sigma_{\rm T}$ is Thomson scattering cross section, $\beta$ is the local fluid velocity in units of light speed $c$,  and $\theta_l$ is the angle between the velocity vector of the local fluid element and the observer's angular location. 

The comoving number density of the plasma moving with outflow rate $\mdj$ is defined as \citep{2013MNRAS.428.2430L},
\be 
n'(r, \theta)=\frac{1}{m_p c \beta \Gamma r^2} \frac{d\mdj}{d\Omega}.
\label{eq_co_den}
\ee 
Here, $m_p$ is proton mass and $d\mdj/d\Omega$ is differential mass outflow rate in the jet.
This differential mass outflow rate is connected to the (differential) jet luminosity ($d\Lj/d\Omega$) as
\be 
\frac{d\mdj}{d\Omega} = \frac{1}{c^2\Gamma}\frac{d\Lj}{d\Omega}.
\ee 
It has been observed that the jet luminosity is not isotropic and has angle dependence. In various observations of extragalactic jets, $L_j \propto \theta^{-2}$ is found to be a typical behaviour of jet luminosity \citep{2001ARep...45..236L,2002ApJ...571..876Z,2002MNRAS.332..945R,2022Galax..10...93S}. In the context of our jet model, we assume that the jet luminosity is constant at small angles, $\theta\le\tj$, and it decreases with inverse square law at larger angles. The angle-independent luminosity at small angles, $\theta\le\tj$, can be understood, since the jet has constant Lorentz factor and hence the flow is radial and steady, leading to angle-independent luminosity. The luminosity can therefore be approximated by  
\be 
\frac{d\Lj}{d\Omega} = \frac{L_0}{2\pi\left[\left(\frac{\theta}{\tj}\right)^2+1\right]} = \frac{L_0}{2 \pi f(\theta)}.
\ee 
Here, $f(\theta)$ is
\be 
f(\theta) = \left(\frac{\theta}{\tj}\right)^2+1.
\ee 
Hence the differential outflow rate in the jet is
\be 
\frac{d\mdj}{d\Omega} = \frac{L_0}{2\pi c^2 \Gamma f(\theta)}.
\label{lab_diff_mdot}
\ee

The total mass outflow rate can be obtained by integrating over the spatial angle $d\Omega$. One can express it in terms of the accretion efficiency in the disk, $\eta=\La/\mda c^2$ where $\mda$ is the disk accretion rate and $\La$ is the accretion luminosity, using
\be 
\mdj = \frac{L_0}{2\pi c^2}\int \frac{d\Omega}{\Gamma f(\theta)}
 = \frac{m_0 L_{\rm a}}{\eta c^2}.
\label{eq_mdot1}
\ee 
Here, $m_0=\mdj/\mda$ is the jet mass loading parameter, typically found between 200-500 \citep{2021ApJ...923..256Q}. As for the accretion efficiency $\eta$, it is found in range $0.016-0.14$ \citep{2003PASJ...55..599B}. Theoretically, it is estimated to be 0.056 for a non rotating black hole while it can be as high as 0.32 in the case of a maximally rotating black hole \citep{1989MNRAS.238..897L}.

The luminosity along the jet axis $L_0$ can be expressed in terms of the disk luminosity, $\La$ as  
\be 
L_0 = \frac{2 \pi m_0 \La}{\eta {\cal{I}}},
\ee
with $\cal{I}$ defined as
$
{\cal{I}} \equiv \int \frac{d\Omega}{\Gamma f(\theta)}.
$ 
Using the expression of $L_0$, Equation \ref{lab_diff_mdot} becomes
\be 
\frac{d\mdj}{d\Omega} = \frac{m_0 \La}{\eta c^2 \Gamma} \frac{1}{f(\theta)\cal{I}},
\ee
%
From Equation \ref{eq_co_den}, the comoving density is 
\be 
n'=\frac{m_0 \La}{\eta m_pc^3\beta \Gamma^2 r^2 f(\theta) \cal{I}}.
\label{eq_nprim}
\ee 
Using this expression in Equation \ref{eq_tau1}, the optical depth can be expressed as
\be 
\tau = \int_{\zmin}^\infty \frac{m_0 \La \sigma_T [1-\beta \cos \theta_l]}{\eta m_pc^3\beta \Gamma r^2f(\theta) {\cal{I}}}dz.
\label{eq_tau3}
\ee 
If the density is $n'=n_0' = n_0/\Gamma_0$ given at distance $r=r_0$ at $\theta = \tj$, then from Equation \ref{eq_nprim},
\be 
\frac{m_0 L_a}{\eta m_p c^3 \cal{I}} = 2 n_0 r_0^2 \beta_0 \Gamma_0.
\label{eq_napp}
\ee 
Here, $\beta_0$ is bulk jet velocity along the jet axis. Using Equation \ref{eq_napp} in \ref{eq_tau3}, 
\be 
\tau = \int_{\zmin}^\infty \frac{2n_0r_0^2\beta_0 \Gamma_0 \sigma_T [1-\beta \cos \theta_l]}{\beta \Gamma r^2f(\theta)}dz.
\label{eq_tau4}
\ee 
The observer, situated at polar angle $\theta_o$, will see the surface of the plasma when the optical depth along the observer's direction equals unity. 
Considering the fact that $r \sin \theta_l$ is constant along the ray parallel to $\theta$, one can convert  the integration from $dz$ to $d\theta_l$ using $dz = -r^2 d\theta/(r\sin \theta_l)$, where $\theta_l$ is the angle between the local propagation direction of flow and the observer's location $\to$. This gives 
\be 
\tau = \int_{0}^{\theta_l} \frac{2n_0r_0^2\beta_0 \Gamma_0 \sigma_T [1-\beta \cos \theta_l]}{\beta \Gamma r \sin \theta_l f(\theta)}d\theta_l.
\label{eq_tau5}
\ee
%
The photons escape to infinity from surface at which $\tau=1$ and that surface is visible to the observer. In general, $\theta_l$ varies along the photon path and is thus $\theta_l \neq \theta_0+\tj$. The surface of the photosphere therefore depends on both the polar and azimuthal location of the observer (\ie $\to$ and $\phi_{\rm o}$).
Equation \ref{eq_tau5} can be solved to determine the photospheric radius for unit optical depth along the direction of the observer $\to$. In general, it is a function of the jet's angular coordinates [\ie $\Rph = \Rph(\theta, \phi)$]. The unique photospheric appearance is determined by specifying the jet parameters according to Equation \ref{eq_gam_profile} and the input parameters given particle density $n'=n_0/\Gamma_0$ at radius $r=r_0$. From various observational estimates, it is found to be in the range of one to few $\times 100$~cm$^{-3}$ at around 1 parsec \citep{1981ApJ...243..700K, 1998A&A...330...79L,2016ApJ...826..135L, 2021ApJ...910...35L}.

\subsection{Numerical simulations}

We carry Monte Carlo simulations to confirm the predictions of the above discussed theoretical model. In the numerical code, approximately $6$ million monoenergetic photons with energy $10^{-10}$ (in units of rest mass of electron) are injected deep inside the jet where the matter is optically thick. Initially the photons have random directions and they propagate while Compton scatter with the electrons in the jet. 

\begin {figure}
\begin{center}
 \includegraphics[width=8cm, angle=0]{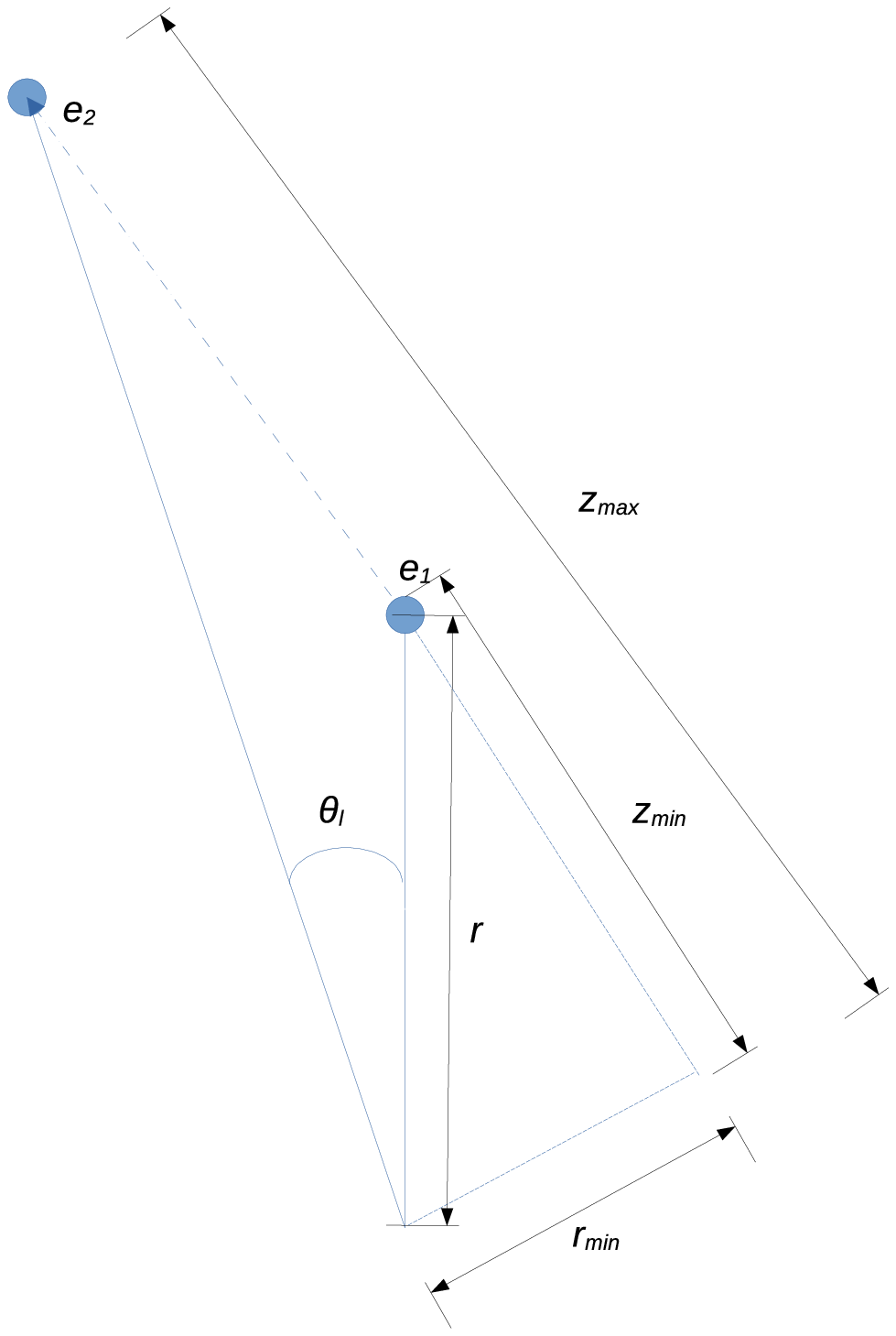}
\caption{Geometry of a photons scattering (dashed line) between the two electrons $e_1$ and $e_2$. The first electron is situated at location $\zmin$ while the second electron at $\zmax$ in cylindrical coordinates chosen such that the z axis is along the photon path. At photospheric radius, $\zmax = \infty$.}
\label{lab_Geom}
 \end{center}
\end{figure}

Given a photon's four vector, the code computes the location of the coming scattering event in the following way. Initially, the photon's location is given in Cartesian coordinate system. As a first step, the photon location is transformed into a cylindrical coordinate system ($r_c, \theta_c, z_c$) . This coordinate system is chosen such that the centre of the black hole is at  $r_c=z_c=0$, and the $z_c$ axis is along the propagation direction of the photon. In these coordinates, the photon's initial location is described by ($\rmin, \zmin$). 
\footnote{As the photon propagates along the $z_c$ direction, $r_{\min}$ and $\theta_c$ are not changed, and we omit discussing $\theta_c$ for clarity.}
The photon's initial radial distance from the center of the black hole is $r = \sqrt{\rmin^2+\zmin^2}$ (see Figure \ref{lab_Geom}).

The photon propagates along the $z_c$ axis (dashed line in Figure \ref{lab_Geom}) until it scatters with an electron $e_2$ located at $z_{\max}$. 
The photon travel distance $\Delta z = z_{\max} - z_{\min}$ is determined as follows. The probability of a photon to travel a distance $ \Delta z$ along which the optical depth is $\tau$ without being scattered  is $P_{sc.}(\tau) = e^{-\tau}$. Therefore, the optical depth $\tau$ is drawn from a logarithmic distribution. Then the code calculates $\delta \tau$ along the photon trajectory until reaching the randomly selected optical depth.  
If the randomly selected optical depth is larger than the optical depth for escaping to infinity, the photon assumes to escape, otherwise the scattering position is calculated along the $z_c$ axis such that the travel distance $\Delta z$ is such that the optical depth along this distance is equal to $\tau$. 
Finally, this scattering location is transformed back to the original Cartesian coordinate system. 

For the scattering process, a cold electron that moves at bulk velocity $\Gamma=\Gamma(\theta)$ as given by Equation \ref{eq_gam_profile} is considered. The scattering occurs at a random angle in the electron's rest frame, considering the full, Klein-Nishina, angle-dependent cross section for the process. For a complete description of the scattering calculations, see \cite{2008ApJ...682..463P}.

A photon scatters multiple times until it escapes the system once it reaches the location where the local optical depth fulfills the escape criteria. The escape direction of each photon is stored and contributes to the observed flux as seen by an observer along the photon escape direction. The escape location marks the position of the photospheric radius perceptible to that observer. This implies that for a photon at the photosphere, $\zmax = \infty$ as the photon has its last scattering there before it escapes to infinity.

Thus, a large number of photons map the entire photospheric surface for all observers. The photospheric surface as seen by observers located in different directions depends on the viewing angle, and is calculated by binning the data into certain observer's locations.

Further details of the simulation code are found in \cite{2008ApJ...682..463P, 2013MNRAS.428.2430L, 2021ApJ...908....9V}.

\begin {figure}
\begin{center}
 \includegraphics[width=8cm, angle=0]{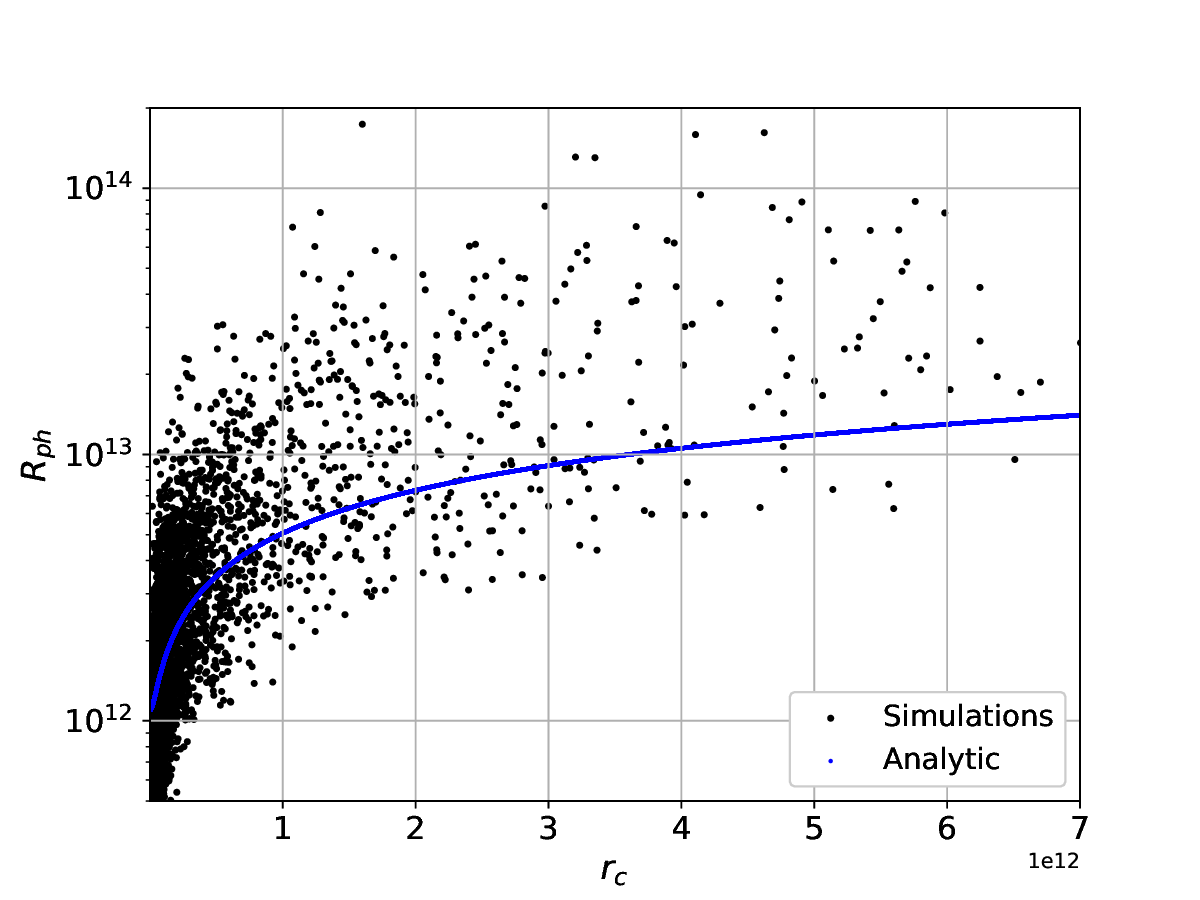}
\caption{$\Rph$ as a function of cylindrical radius $r_c = \sqrt{x^2+y^2}$ with $\Gamma_0 = 20$, $\tj=0.1$ rad, $p=1.2$, $\to=0.0$ rad for both analytic calculations (blue dots) and results from Monte Carlo simulations (Black spheres). $n_0=80$ cm$^{-3}$ at $r_0 = 1$ parsec. 
}
\label{lab_compare_simulations_theory}
 \end{center}
\end{figure}

\begin {figure}
\begin{center}
 \includegraphics[width=8cm, angle=0]{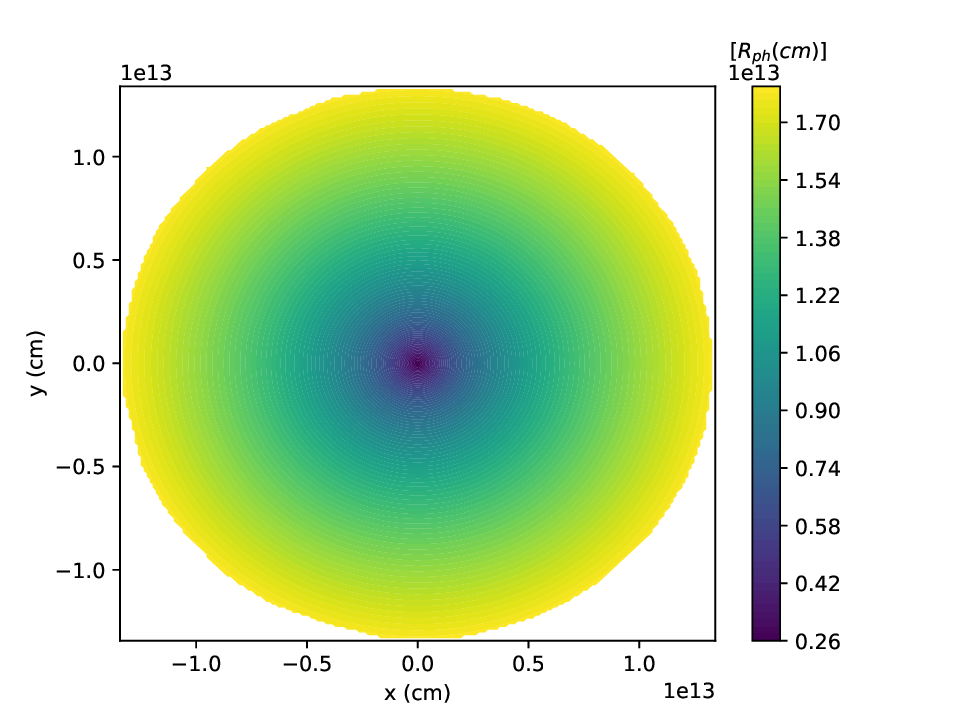}
 \includegraphics[width=8cm, angle=0]{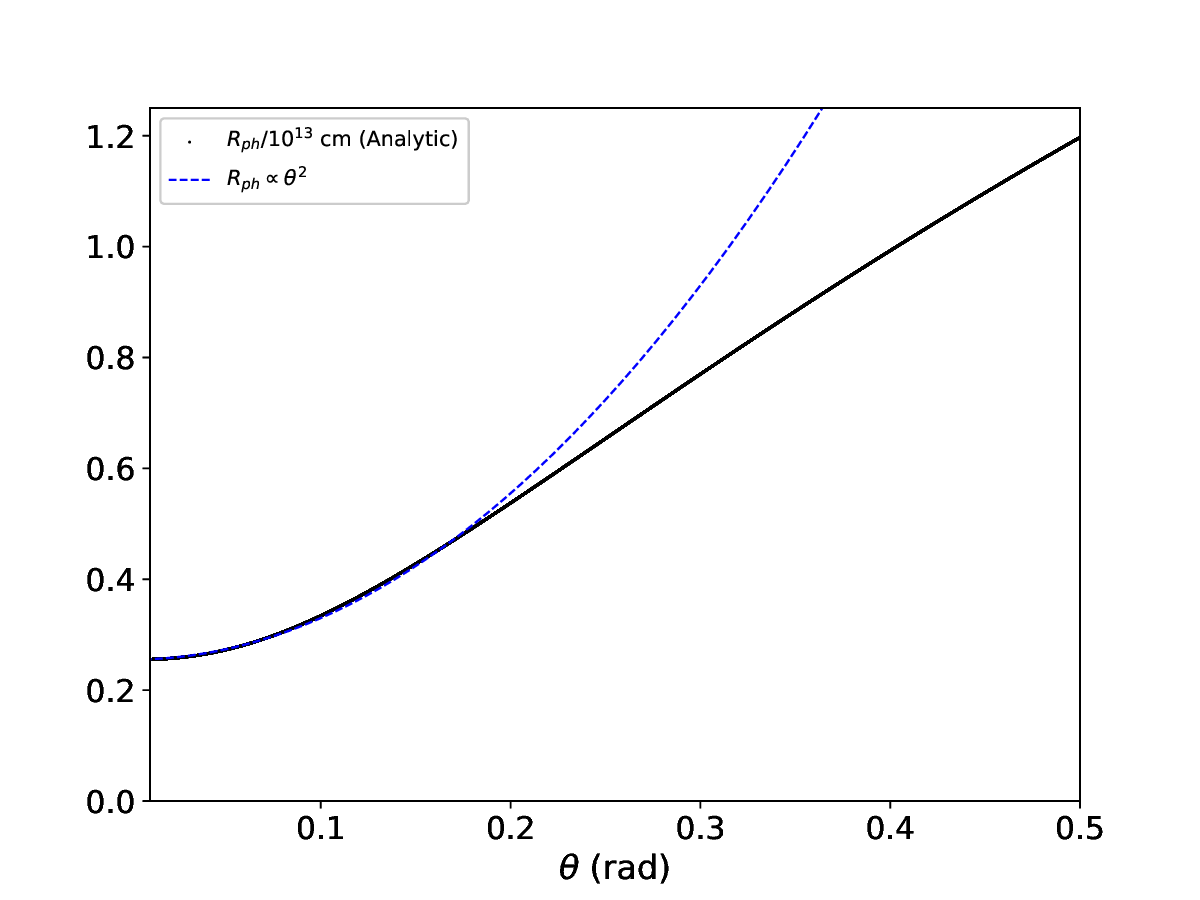}
\caption{Upper Panel: $\Rph$ as a function of $x$ and $y$ coordinates (in cm). Lower panel: $\Rph$ as a function of the polar angle $\theta$. The color gradient shows the luminosity of the photosphere decreasing from blue to yellow shade. Chosen parameters $\Gamma_0 = 10$, $\tj=0.1$ rad, $p=1.2$ for the observer's location at $\to=0.0$ rad. $n_0=80$ cm$^{-3}$ at $r_0 = 1$ parsec}
\label{lab_g10tj0.1p1.2tho0}
 \end{center}
\end{figure}

\section{Results}
\label{sec_results} 
As an example, we consider an environment characterized by a particle density $n_0 = 80$ cm$^{-3}$ at $r_0 = 1$ parsec. In Figure \ref{lab_compare_simulations_theory}, we plot the photospheric radius $\Rph$ with horizontal distance from the jet axis $r_c = \sqrt{x^2+y^2}$.  The observer is assumed along the jet axis (polar angle $\to=0$). The estimated values of $\Rph$ from the analytic calculations (Equation \ref{eq_tau5}) are shown by blue curve while the inferred values from Monte Carlo simulations are overplotted with black dots. The edge brightening of the jet due to relativistic effects on the apparent optical depth is perceptible as the photospheric radius extends up to larger distances for off axis regions compared to the jet axis. Considering a black hole mass equivalent to $10$ million solar mass, (with Schwarzschild radius $\sim 3 \times 10^{12}$~cm), the photosphere is well above the horizon, extending up to one order of magnitude above it. However, the extent of the magnitude primarily depends upon considered value of $n_0$ and could extend much further for larger densities.


\begin{figure}
\begin{center}
 \includegraphics[width=9cm, angle=0]{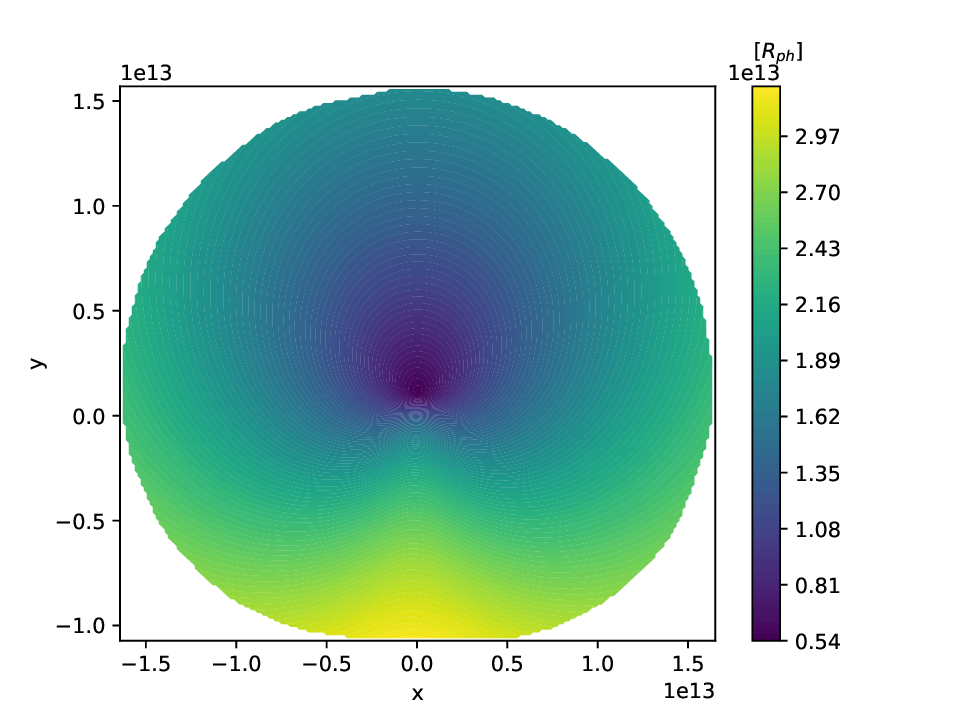}
  \includegraphics[width=9cm, angle=0]{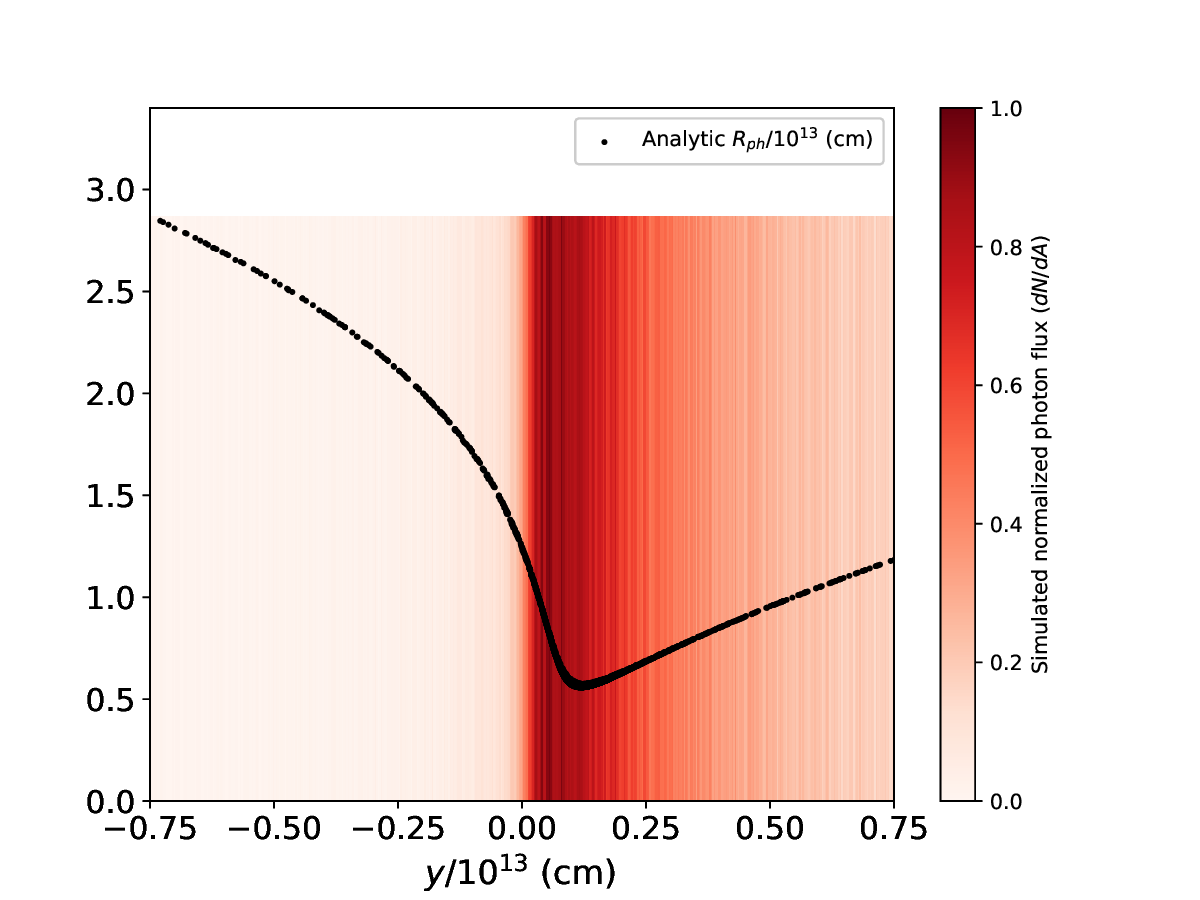}
  \includegraphics[width=9cm, angle=0]{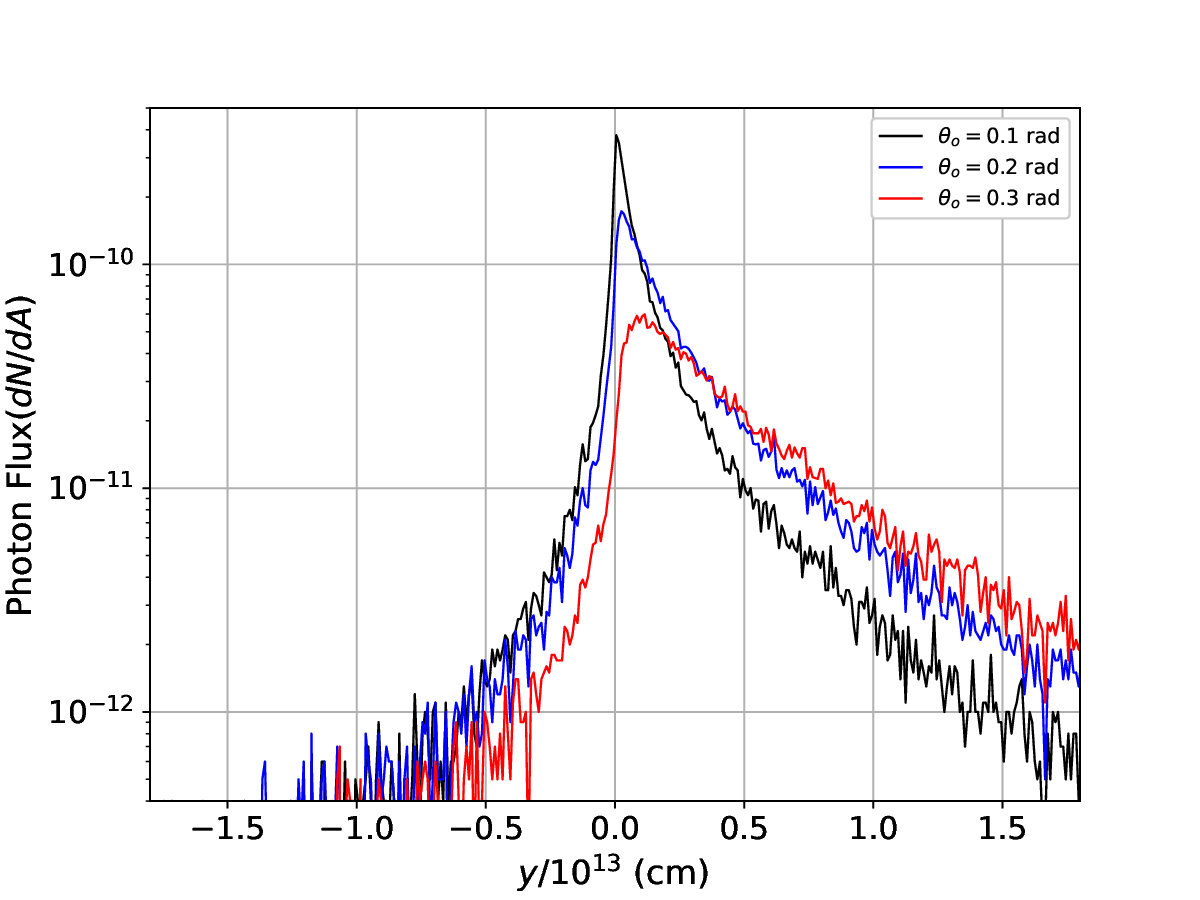}
\caption{Top panel: Analytic estimation of $\Rph$ as a function of $x$ and $y$ coordinates for $\to=0.3$ rad; Middle panel: for $\to=0.3$ rad, we plot the variation of $\Rph$ along the y axis for $x=0$ with black spheres. The colored plot is the normalized photon flux $dN/dA$ calculated by simulations at a given location $x=0, y$. Bottom panel: variation of the simulated photon flux along the $y$ coordinate at $x=0$ for different observing angles $\to=0.1$ rad (black), $\to=0.2$ rad (blue) and $\to=0.3$ rad (red) curves. For larger viewing angles the asymmetry is larger. Chosen parameters in all the panels are $\Gamma_0 = 10$, $\tj=0.1$ rad, $p=1.2$. $n_0=80$ cm$^{-3}$ at $r_0 = 1$ parsec.}
\label{lab_sp_theta_p0.30}
 \end{center}
\end{figure}


To show the appearance of the jet to an observer, analytic results are shown in Figure \ref{lab_g10tj0.1p1.2tho0} (upper panel) where we project the photospheric radius $\Rph$ as measured in Cartesian coordinates on the $x-y$ plane. Here, the observer is assumed on axis, namely situated along the $z$ axis (at polar angle $\to = 0$). 
This makes the surface symmetric around the $x$ and $y$ coordinates. In producing this plot, we have assumed a jet profile as given in Equation \ref{eq_gam_profile}, with parameters $\Gamma_0 = 10$, $\tj=0.1$ rad and profile index $p=1.2$.

The important outcome is the increase of $\Rph$ at the jet's edges up to several times compared to the jet axis (at $x=y=0$). The blue region at the centre of the jet compared to yellow region at outskirts signifies a deeper photosphere at the centre. 
In the lower panel, we have plotted the photospheric radius $\Rph$ as a function of the polar angle $\theta$ (black solid curve). In the inner regions of the jet, the photospheric radius $\Rph$ is well approximated as a parabolic function of the polar angle, obeying the relation $\Rph = a\theta^2+b$ where $a=5\times 10^{13}$ cm and $b=2.55\times 10^{12}$ cm for the jet profile presented here (blue dashed curve). 

In the upper panel of figure \ref{lab_sp_theta_p0.30}, we show the photospheric radius as seen by an off-axis observer. The photospheric radius $\Rph$ is plotted on the $x-y$ plane for jet parameters $\Gamma_0 = 15$, $\tj=0.1$ rad, $p=1.2$. The observer assumes $\to = 0.30~$rad and an azimuthal location $\phi_{\rm o} = \pi/2$, making the surface symmetric around the $x$ axis, while asymmetry appears along $y$ axis. 
To indicate the observed asymmetry of the photospheric radius and observed flux for such an off-axis observer, we plot the photospheric radius $\Rph$ along the $y$ axis in the middle panel of figure \ref{lab_sp_theta_p0.30} for $x=0$ (black dots). The color bar shows the simulated photon flux $dN/dA$, where $dA(y)=dxdy$ is the differential cross section shown along the $y$ axis. 
The base of the photospheric surface (minimum of the last scattering surface) shifts towards positive $y$ axis. The bright spot peaks near the dip (the smallest photospheric radius, at $5 \times 10^{12}$ cm). It asymmetrically extends towards both sides. The arm along the left side (further away from the observer) appears twice as long compared to right side, which is purely an effect of the observer's location. Such asymmetry of the limb brightened arms is widely observed in off axis jets \citep{2016Galax...4...46W}. The shorter arm appears brighter compared to the longer one. In the bottom panel, we show the variation of the simulated photon flux along $y$ coordinate at $x=0$ for different choices of the observer's location at $\to=0.1,0.2$ and $0.3$ rad. The asymmetry in the flux decay at both sides increases as the observer is further away from the jet axis. Subsequently, the peak in the flux also shifts further along $y$ axis.
Further, the implication of this asymmetry in the flux is that by looking at the ratio of the observed signal at both sides, one can infer the viewing angle of the jet. 


\section{Conclusions}
\label{sec_conclusions}
In this letter, we have studied the photospheric appearance of a relativistic AGN jet with polar angle dependent Lorentz factor profile. Using density estimates of AGN jets, we argue that the jet cores near their launching sites, and up to a few- few tens of Schwarzschild radii can generally be optically thick for Compton scattering. Photons undergo multiple scattering with electrons in this region, before they escape from it to infinity. The last scattering surface represents the innermost region that can be observed. 

Here we considered an angle dependent Lorentz factor profile, with jet propagating faster in the middle compared to larger polar angles. We have further considered a cold plasma. However, the last scattering surface, or the structure of photosphere obtained in this work remains unaffected if the plasma were hot and relativistic. Furthermore, as discussed in section \ref{sec_sub_compton_phot}, the optical depths at the jet base are around $10$ and hence the photons decouple from the plasma after encountering only a few scatterings before escaping. Thus, complete thermalization does not have time to occur and most of the photon population remains at low energies, enabling the resolved cores to be observed at radio frequencies.

We show that for an observer situated near the jet axis, the jet core appears relatively empty in the middle with a deeper photosphere while it appears extended at the edges. The reason of this appearance is that the optical depth for scattering is smaller for photons emitted along the jet axis, compared to photons emitted from larger angles. As a result, the last scattering location of the photons is closer to the jet base along the jet axis.

Such an appearance is consistent with various observations of the resolved jet cores. Hence, we provide a natural explanation to the observed edge brightening feature of the AGN jet cores. 
As the appearance of a jet base depends upon jet parameters, the analysis hands over a tool to infer the jet structure in AGNs directly from their observations. 

The conventional explanation of the limb brightening due to Doppler deboosting model is criticized based on the fact that it cannot explain persisting limb brightening in a swinging jet. Thus, when the jet points towards the observer, the limb brightening should vanish which is in contrast with observations of Mrk 501 \citep{2021Galax...9...58G}. From Figure \ref{lab_g10tj0.1p1.2tho0} we show that our model predicts a limb brightening for an on axis jet as well. Thus it is a viable mechanism that explains different aspects of the limb brightening phenomenon, that are difficult to explain otherwise. 

The relative flux variation at both sides of the jet depends upon the observer's location as shown in the bottom panel of Figure \ref{lab_sp_theta_p0.30}. This enables the determination of the observer's location from the observed asymmetry in the flux variation. Additionally,  the flux decays by two to three orders of magnitude between the centre of the jet and the off axis regions or the limb. This is consistent with the typically observed flux variation in the resolved cores of AGN jets \cite[See Figure 1 of ][]{2013ApJ...775...70H}. 

The analysis presented here paves a way for considering the presence of a photospheric component in blazars. The existence of an observed photospheric component may have additional, exciting implications on the observed AGN properties, such as the spectra and polarization. These will be explored in future works. 

\section*{Acknowledgments}
For this project, we acknowledge the support by European Union (EU) via ERC consolidator grant 773062 (O.M.J.).

\bibliography{ref1}{}
\bibliographystyle{aasjournal}
\end{document}